\documentclass[journal=esthag,manuscript=article,layout=onecolumn]{achemso}
\usepackage{natbib}
\usepackage{chemformula} 
\usepackage[T1]{fontenc}
\usepackage{amsmath}
\usepackage{chemformula}
\usepackage{subcaption}
\usepackage{bm}
\pdfoutput=1 
\usepackage{lineno}

\usepackage{hyperref}
\SectionNumbersOn

\DeclareUnicodeCharacter{2009}{\,} 
\DeclareUnicodeCharacter{2082}{$_2$}
\DeclareUnicodeCharacter{2212}{-} 

\author{Ishaan Markale}

\alsoaffiliation[EAWAG]
{Eawag, Swiss Federal Institute of Aquatic Science and Technology, 8600 Dübendorf, Switzerland}
\affiliation[ETH Zürich]
{Department of Civil, Environmental and Geomatic Engineering, ETH Zürich, Stefano-Franscini-Platz 5, 8093 Zürich, Switzerland}
\author{Gabriele M. Cimmarusti}
\affiliation[University of Birmingham]
{School of Chemistry, University of Birmingham, Edgbaston, Birmingham, B15 2TT, UK}
\author{Melanie M. Britton}
\affiliation[University of Birmingham]
{School of Chemistry, University of Birmingham, Edgbaston, Birmingham, B15 2TT, UK}
\author{Joaqu{\'{i}}n Jim{\'{e}}nez-Mart{\'{i}}nez}
 \email{joaquin.jimenez@eawag.ch / jjimenez@ethz.ch}
\alsoaffiliation[EAWAG]
{Eawag, Swiss Federal Institute of Aquatic Science and Technology, 8600 Dübendorf, Switzerland}
\affiliation[ETH Zürich]
{Department of Civil, Environmental and Geomatic Engineering, ETH Zürich, Stefano-Franscini-Platz 5, 8093 Zürich, Switzerland}

\title{Phases saturation control on mixing driven reactions in 3D porous media}

\keywords{mixing-driven reaction, MRI, phase saturation, 3D pore scale reactions, reactive lamella model}

\begin{document}

%%%%%%%%%%%%%%%%%%%%%%%%%%%%%%%%%%%%%%%%%%%%%%%%%%%%%%%%%%%%%%%%%%%%%
%% The "tocentry" environment can be used to create an entry for the
%% graphical table of contents. It is given here as some journals
%% require that it is printed as part of the abstract page. It will
%% be automatically moved as appropriate.
%%%%%%%%%%%%%%%%%%%%%%%%%%%%%%%%%%%%%%%%%%%%%%%%%%%%%%%%%%%%%%%%%%%%%

\begin{figure*}
\begin{tocentry}
	\centering
	\includegraphics[width = 5.5cm]{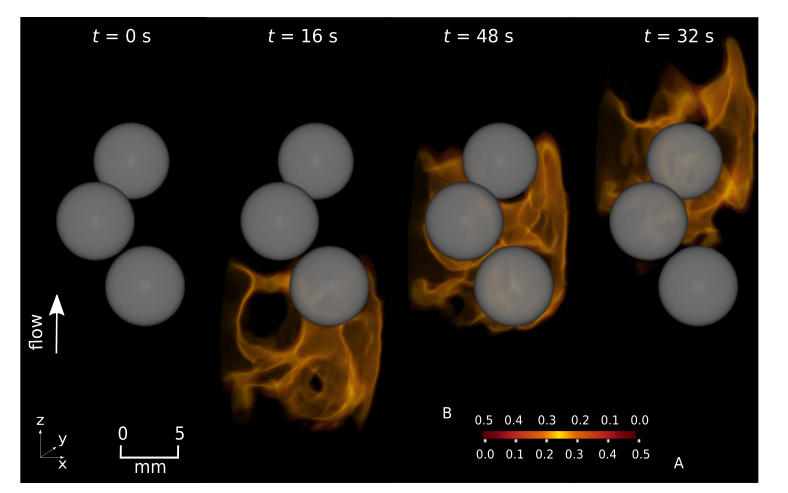}
	\label{fig:TOC}
\end{tocentry}
\end{figure*}

%%%%%%%%%%%%%%%%%%%%%%%%%%%%%%%%%%%%%%%%%%%%%%%%%%%%%%%%%%%%%%%%%%%%%
%% The abstract environment will automatically gobble the contents
%% if an abstract is not used by the target journal.
%%%%%%%%%%%%%%%%%%%%%%%%%%%%%%%%%%%%%%%%%%%%%%%%%%%%%%%%%%%%%%%%%%%%%
\begin{abstract}

Transported chemical reactions in unsaturated porous media are relevant across a range of environmental and industrial applications. Continuum scale dispersive models are often based on equivalent parameters derived from analogy with saturated conditions, and cannot appropriately account for processes such as incomplete mixing. It is also unclear how the third dimension controls mixing and reactions in unsaturated conditions. We obtain 3$D$ experimental images of the phases distribution and of transported chemical reaction by Magnetic Resonance Imaging (MRI) using an immiscible non-wetting liquid as a second phase and a fast irreversible bimolecular reaction. Keeping the Péclet number (Pe) constant, we study the impact of phases saturation on the dynamics of mixing and the reaction front. By measuring the local concentration of the reaction product, we quantify temporally resolved effective reaction rate ($R$). We describe the temporal evolution of $R$ using the lamellar theory of mixing, which explains faster than Fickian ($t^{0.5}$) rate of product formation by accounting for the deformation of mixing interface between the two reacting fluids. For a given Pe, although stretching and folding of the reactive front are enhanced as saturation decreases, enhancing the product formation, this is larger as saturation increases, i.e., volume controlled. After breakthrough, the extinction of the reaction takes longer as saturation decreases because of the larger non-mixed volume behind the front. These results are the basis for a general model to better predict reactive transport in unsaturated porous media not achievable by the current continuum paradigm.
\end{abstract}

\section{Introduction}

The chemical and biological evolution of many natural, engineering and industrial systems is governed by reactive mixing interfaces \cite{Simoni2005,Tartakovsky2008,Rezaei2005,Ottino1989}. Chemical reactions in fluid are driven by mixing, which is a process that brings together segregated substances. Mixing is primarily driven by stretching, due to the existence of fluid velocity gradients, and diffusion \cite{Ranz1979,Villermaux2003,Duplat2008}. Porous media are topologically complex environments with highly heterogeneous fluid flow dynamics. Under unsaturated conditions, i.e., in the presence within the pore space of another phase such as an immiscible liquid or gas, this is exacerbated by a much more complex spatial configuration of the phases. This affects the internal connectivity of the system \cite{JJMNegre2017}, further increasing the fluid flow heterogeneity with the formation of preferential paths (high velocity zones) and stagnation zones (low velocity zones). Understanding mixing and reactions in unsaturated flows in porous media is fundamental to predicting the dynamics of contaminants and to evaluating the role of the soils in controlling global carbon, i.e., soil respiration, \cite{Xu2004,Giardina2014,Ebrahimi2018} nitrogen \cite{Sebilo2013,Helton2015,Kravchenko2017}, and trace element \cite{Winkel2015} cycles. It is also key as it plays a major role in applications such as catalytic reactors \cite{dudukovic2002multiphase}, contaminant (bio-)remediation \cite{Rolle2009}, enhanced oil recovery \cite{JJM2016} and nuclear waste disposal \cite{Winograd1981}.

The classic modeling approaches, also called Fickian or dispersive-diffusive, for reactive transport in unsaturated conditions are often based on equivalent parameters (e.g., dispersion) derived from analogy with saturated conditions, with systematic \textit{ad hoc} incorporation of saturation dependency (fraction of the pore volume occupied by one of the immiscible phases) \cite{Simunek2008}. Fundamentally, these models cannot predict accurately the kinetics of transported reactions resulting from mixing, which intrinsically occur at pore scale \cite{Rolle2009,Williams2009}. Stretching and folding of reactive fronts by the high flow heterogeneity enhance mixing and thus reaction compared to diffusive mixing only \cite{Heyman2020,Wright2017,LeBorgne2014}. Recently developed lamellar mixing models that couple stretching and diffusion to capture the pore-scale concentration fluctuations are a promising avenue to predict reactive processes in these highly complex systems \cite{LeBorgne2013,LeBorgne2015,Lester2016}. 

Mixing and reactive transport at pore scale has been studied experimentally using 2$D$ milli- and microfluidic approaches in saturated and unsaturated conditions \cite{DeAnna2013,JJM2015,JJM2017,Karadimitrou2016}. Various numerical studies have further added insights under both conditions \cite{Willingham2008,Li2019,JJM2020}. However, the impact of incomplete mixing processes and non-Fickian dispersion on reaction kinetics is still not completely understood, especially in 3$D$ \cite{Dentz2011}, where the third dimension adds spatial heterogeneities to the system and greater tortuosities \cite{Ghanbarian2013}. While some recent studies, both numerical and experimental, have addressed the complexity added by the third dimension on mixing and reactive processes \cite{Comolli2019,Heyman2020}, our current understanding of flow dynamics and associated reactive processes in unsaturated porous media is very limited, owing to both the complexities associated with the presence of multiple phases and the difficulty of experimentally (in particular, optically) accessing these systems.
 
The development, over the last few decades, of non-invasive and non-destructive 3$D$ imaging techniques at pore scale opens a wide spectrum of possibilities to address the complexity of natural media, in which accessibility is limited \cite{Britton2005,Wildenschild2002,Krummel2013,Heyman2020}. To date, the use of techniques such as laser tomography, confocal microscopy, X-ray absorption computed tomography or magnetic resonance imaging (MRI) has been limited to imaging the distribution of phases and multiphase flow \cite{Berg2013}, fluid flow in saturated and unsaturated media \cite{Deurer2002,Krummel2013}, conservative transport in saturated conditions \cite{Greiner1997,Heyman2020}, and propagation of reactive waves\cite{Britton2005,Rose2013}. In this work, we employ MRI to visualize the 3$D$ distribution of phases and reactants in unsaturated porous media.

The main goals of the present work are to measure pore-scale dynamics of local concentration to identify the mechanisms that control kinetics of transported reactions in unsaturated porous media, and to systematically characterize the impact of phase saturation on the effective reaction rate of the system, i.e., on the product formation. For this purpose, we use an analogous porous medium consisting of glass beads and a catalyst/indicator reaction (mixing-limited bimolecular irreversible reaction). Two different glass bead sizes are used for the analysis, and several phases saturation and flow rates are explored. This experimental approach provides a completely new perspective for reactive transport in unsaturated porous media, and the experimental results are used to develop a theoretical framework in 3$D$ based on the lamella mixing and reaction models by providing scaling laws for the effective reaction rate as a first step in the linking of the pore-scale phenomena to the Darcy (continuum) scale.

\section{Reactive Lamella Theory} \label{section:theory}
%Porous media (consisting of sand, pore size 50 µm to 2 mm; silt, pore size 2–50 µm; and clay, pore size < 2 µm).

\subsection{Transport and Mixing}

The transport at pore scale of two initially segregated reactants is governed by advection, dispersion and reaction. In the absence of inertia effects, Navier-Stokes equation can be used as the governing equation:

\begin{equation}
	\label{eqn:ADR}
	\frac{dc_\textrm{i}}{dt} + \nabla \cdot(\bm{v} c_\textrm{i}) - D\nabla^2 c_\textrm{i}= r_\textrm{i}
\end{equation}

where $c_\textrm{i}$ is the concentration of the respective reactants, $D$ is the molecular diffusion coefficient, $\bm{v}$ is the velocity field (calculated from the solution of the flow problem), and $r_\textrm{i}$ is the local reaction rate. The heterogeneous velocity field, with complex streamline topologies, leads to the deformation of the interface, i.e., mixing front, between reactants. The deformation of the mixing front can enhance mixing and reaction rates by increasing the area available for diffusive mass transport \cite{LeBorgne2014,JJM2015,JJM2017}.

We consider an initially 3$D$ flat (non-deformed) interface involving a mixing driven irreversible reaction $A+B\rightarrow P$ in a porous medium. The lamella theory of mixing and reaction, a Lagrangian framework that links the distribution of stretching rates along mixing interfaces to mixing and reaction rates \cite{LeBorgne2014,LeBorgne2013,LeBorgne2015}, assumes that the mixing front is a collection of numerous stretched lamellae, also called sheets in 3$D$ \cite{Ruiz2018}. At pore scale, the deformation of the mixing interface of area $\varepsilon$, with initial area $\varepsilon_0$, is quantified by the elongation $\rho = \varepsilon/\varepsilon_0$ \cite{Meunier2010} and its transport as \cite{Ranz1979,Bando2017,Bando2018}:

\begin{equation}
	\label{eqn:ADR_lamella}
	\frac{dc_\textrm{i}}{dt} - \gamma n \frac{dc_\textrm{i}}{dt} - \frac{d}{dn} \left(D\frac{dc_\textrm{i}}{dn}\right) = r_\textrm{i} 
\end{equation}
where $\gamma$ is the stretching rate defined as $\gamma = (1/\rho)(d\rho/dt)$ and $n$ is the coordinate perpendicular to the lamella. Since the concentration gradients along lamella are small \cite{LeBorgne2014}, we assume that diffusive mass transfer is dominant only perpendicular to the lamella, i.e., along $n$.

\begin{figure}
	\centering
    \includegraphics[width=7.5cm]{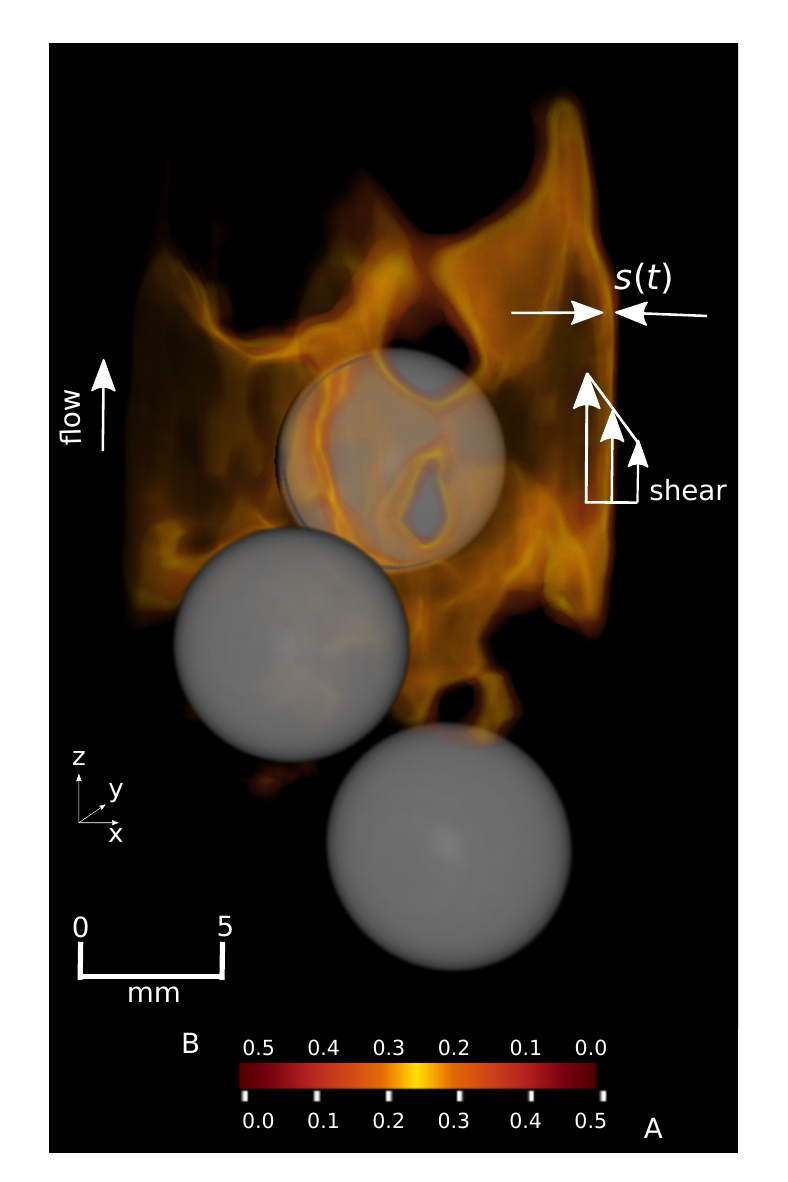}
	\caption{3$D$ reactive lamella inside a fully saturated ($S_{\textrm{w}}=1$) packed bed of 8 mm glass beads (only three beads in grey color are shown for visualization simplicity). Concentration of the reactants within the mixing volume is shown in warm colors, in which the lightest color indicates equal concentration of the invading ($A$) and resident ($B$) reactant. $A$ is pumped into the porous medium at constant flow rate ($Q=0.252$ mm$^3$/s) from bottom to top. Shear is indicated with velocity vectors of different magnitude. The width or transverse thickness of the lamella is labeled as $s(t)$. Note that the lamella shown is cut by an x-z (vertical) plane passing through the middle of the domain; beads are not cut by the plane (see SI, Movie S1).}
	\label{fig:8mm}
\end{figure}

The transport and reaction regimes are characterized by the dimensionless numbers Péclet (Pe) and Damköhler (Da), respectively. Péclet number, $\textrm{Pe} = \tau_{\textrm{d}} / \tau_{\textrm{a}} = \bar{v}\xi/2D$, represents the ratio between the characteristic time of diffusion and the characteristic time of advection over a typical pore throat $\xi$, being $\bar{v}$ the mean pore water velocity. Damköhler number, $\textrm{Da} = \tau_{\textrm{a}}/ \tau_{\textrm{r}}= \xi c_0 k/\bar{v}$, represents the ratio of transport to reaction timescales. The characteristic reaction time, under well mixed conditions, is calculated as $\tau_{\textrm{r}} =  1/c_0k$, where $k$ is the rate constant of our chosen reaction and $c_0$ is the initial concentration. When reaction time scale is much smaller than the advection and diffusion time scale, i.e., $\textrm{Da}>>1$, the reaction is driven by mixing.

\subsection{Effective Reaction Rate: Mass of Product}

For a fast irreversible reaction, in which the mass of $A$ that diffuses from the interface into $B$ reacts to produce $P$, the global kinetics, i.e., effective reaction rate $R$, of mass of product $M_P$ over the interface $\Pi$ can be written as \cite{DeAnna2013,Bando2017}:

\begin{equation}
	\label{eqn:model}
	R=\frac{dM_P}{dt} = D \int_{\Pi}|\nabla c_A|d\varepsilon \approx \frac{D c_0  S_\textrm{w}\varepsilon_0 (1 + \gamma t)}{s(t)}
\end{equation}

where $s$ is the width or transverse thickness of the interface (see Figure \ref{fig:8mm}). It is assumed that $|\overline{\nabla c}|\sim c_0/ s$. In natural systems, and in unsaturated soils in particular (with pore sizes [1-10$^{-4}$] mm and pore flow velocities [10$^{-2}$-10$^{-6}$] mm/s) \cite{Vanderborght2007}, a shear flow regime is expected, in which the mixing front deforms by the gradient of velocity in the direction transverse to the main flow \cite{LeBorgne2015}. As detailed for 2$D$ flows \cite{deAnna2014_2,Bando2018}, shear flows in 3$D$ can also lead to a linear increase of elongation ($\rho= \nabla v t$) and a linear increase of the mixing interface area as $\varepsilon = S_\textrm{w}\varepsilon_0 (1+\gamma t)$, \cite{Ruiz2018} where $S_\textrm{w}$ is the wetting saturation defined as ratio of the volume occupied by the wetting phase (i.e., phase in which reaction is taking place) to that of the porous space. We estimate $\gamma$ as $\overline{v}/\lambda = \overline{v}/(\xi/S_\mathrm{w})$, where $\lambda$ is the velocity correlation length \cite{Villermaux2012}. $\lambda$ increases as $S_\mathrm{w}$ reduces as has been observed recently \cite{Velasquez2020,An2020}. The elongation of the mixing front increases the area available for diffusive mass transport, but also enhances the concentration gradients by compression normal to the surface. The competition between compression and molecular diffusion controls $s$ as follow: 

\begin{equation}\label{eqn:s}
	s(t) = s_0 \sqrt{\frac{3\beta -2 + 2(1+ \gamma t)^3}{3\beta (1+\gamma t)^2}}
\end{equation}
where $\beta = s_0^2\gamma/D$ and $s_0$ is the initial interface thickness. When compression balances with diffusion, concentration gradients decrease, and $s$ grows diffusively as $s \sim t^{1/2}$. The time at which concentration gradients are maximum corresponds to the so-called mixing time $t_{\textrm{mix}}  = \nabla v^{-1} (s_0^2\nabla v /D)^{1/3}$. \cite{Villermaux2012,Bando2018,LeBorgne2015} Note that the characteristic shear time, defined as $\tau_{\textrm{s}}=\nabla v^{-1}$, and the characteristic reaction time $\tau_{\textrm{r}}$ have been used to define the transition between different temporal scaling laws for $R$. Analytical expressions for those scaling laws have been derived for both weak (Pe'< Da') and strong  (Pe' > Da') stretching, with $\textrm{Pe'} = \tau_{\textrm{D}} / \tau_{\textrm{s}}$ and  $\textrm{Da'} = \tau_{\textrm{D}}/ \tau_{\textrm{r}}$, and where $\tau_{\textrm{D}}=s_0^2/D$.\cite{Bando2017} The different temporal scaling laws for the reaction rate $R$ controls the mass of product $M_P$. The mass is computed as $M_P(t)=\int_{V}c_Pd\textbf{x}$ at the instant $t$, where $V$ is the total pore volume of the phase in which transport and reaction are happening.

\subsection{Incomplete Mixing Behind the Front}

The incomplete mixing behind the reactive front makes the reaction persists locally in space and longer in time. The existence of concentration gradients mainly transverse to the main flow direction are responsible of the resilience in the system \cite{Meunier2003}. The decay of the concentration gradients and therefore of the reaction is mainly driven by the diffusion into the low velocity regions or stagnation zones \cite{JJM2015,JJM2017}. For a given position, i.e., a plane transverse to the main flow direction, after breakthrough of the reactive front, the reaction rate $R$ drops and can be scaled by a power law with an exponent, $\alpha$ \cite{Dentz2018,Nissan2020}. This scaling is expected to be controlled by saturation $S_{\textrm{w}}$, but also by Pe and Da.

\section{Materials and Methods} \label{section:methods}

\subsection{Porous Medium: Experimental Design} 

\begin{figure} %[h]
 	\centering
 	\includegraphics[width=8cm]{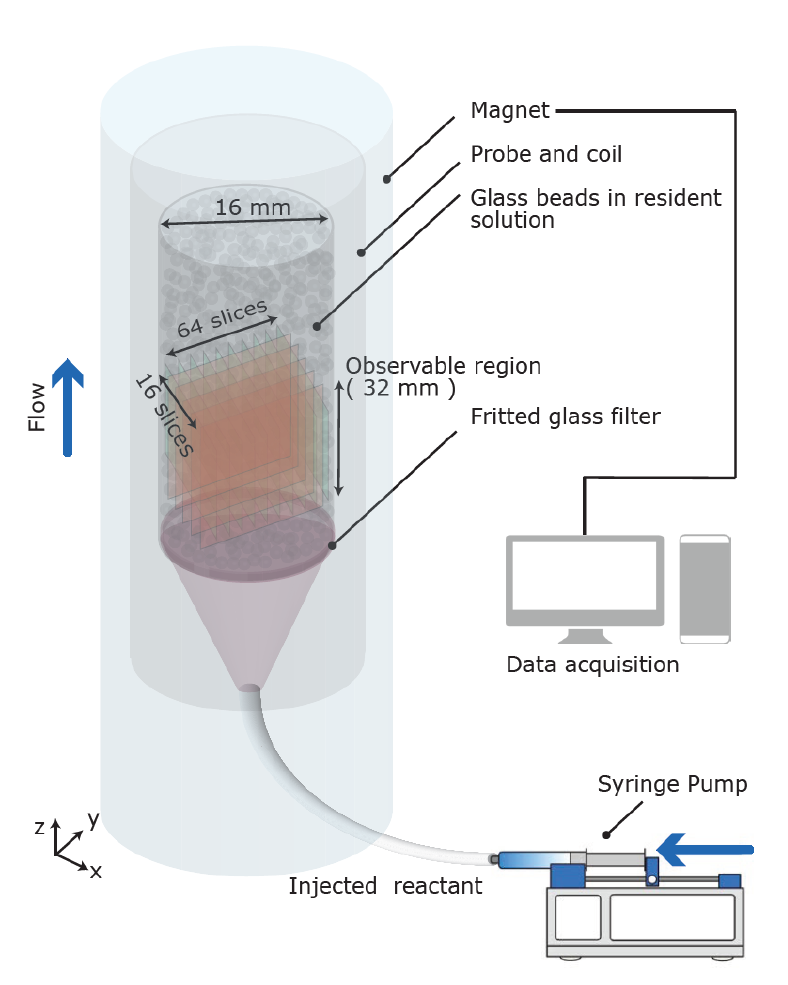}
 	\caption[Experimental setup]{Scheme of the experimental setup. Glass beads are held within a glass cylinder by a fritted glass filter located at the bottom of the column. A syringe pump is used to inject reactant $A$ at constant flow rate into the column containing glass beads and reactant $B$. Flow sense is upward. The immiscible phase ($O$) for unsaturated experiments is allocated within the porous medium accessing it from the upper part of the column. The porous medium column is placed inside the MRI magnet used for imaging the reaction. The visible field of view is 32 $\times$ 16 mm (vertical $\times$ horizontal).}
 	\label{fig:schematic}
\end{figure}

In this work we consider a monodispersed 3$D$ porous medium consisting of spherical grains (glass beads) contained in a cylindrical column of internal diameter 16 mm. Two different glass bead diameters (1 and 4 mm) are used, allowing to explore a wide range of experimental conditions. The column consists of a conical section at the bottom, a diffusing section, and a cylindrical section at the top, which represents our porous medium. Through the conical section, the injected liquid transitions from the diameter of the connecting pipe to the diameter of the cylindrical section. The height of the cone was designed with the help of computational fluid dynamics to minimize the boundary effects for the experimental flow rates expected (see SI, Figure S1). The diffusing section, composed of a fritted glass filter of 2 mm thickness and pore size  500 µm, is used to hold the beads and to homogenize the flow of the invading reactant liquid (henceforth called $A$). It also acts as the limit between the invading and resident (henceforth called $B$) reactant during the experiment set-up. The column is connected to a syringe pump (Harvard Apparatus PHD Ultra) to control the injection flow rate ($Q$) of the injected reactant $A$. We control the syringe pump remotely using the FlowControl software \cite{FlowControl_software} (Figure \ref{fig:schematic}).

As immiscible non-wetting phase, a fluorinated hydrocarbon, Tetrafluorohexane ($\mathrm{C_6F_{14}}$, henceforth called $O$) is used for the unsaturated experiments. Tetrafluorohexane is a stable organic compound which does not react with any of the chemicals used (see Section \ref{section:MRI}) and is not miscible with water. 

\subsection{MRI: From Signal Intensity to Concentration} \label{section:MRI}

We chose a manganese reduction reaction to study the transport of reaction front. A resident solution of Mn\textsuperscript{3+} ions ($B$) in the pore space is displaced by an invading solution containing Mn\textsuperscript{2+} ($A$) and the reactant CHD. The reduction reaction consists of two steps (see Eq. \ref{eq:reaction}) and a conversion from Mn\textsuperscript{3+} to Mn\textsuperscript{2+} \cite{Britton2006_2}. The reacting solution used is H\textsubscript{2}SO\textsubscript{4} (Sigma Aldrich), 1,4-cyclohexanedione (CHD, Sigma Aldrich) and manganese(III) acetate (Sigma Aldrich): 

%\begin{align}
%	\mathrm{2Mn^{3+} + CHD \rightarrow 2Mn^{2+} + H_2Q + 2H^+} \nonumber \\
%	\mathrm{2Mn^{3+} + H_2Q \rightarrow 2Mn^{2+} + Q + 2H^+}\label{reaction}
%\end{align}

\begin{subequations}
	\label{eq:reaction}
	\begin{align}
		\mathrm{2Mn^{3+} + CHD \rightarrow 2Mn^{2+} + H_2Q + 2H^+} \label{eq:reaction_1}\\
		\mathrm{2Mn^{3+} + H_2Q \rightarrow 2Mn^{2+} + Q + 2H^+}\label{eq:reaction_2}
	\end{align}
\end{subequations}

where H$_2$Q is an intermediate organic molecule, 1,4-hydroquinone, and Q is quinone. A solution of $0.5 \times 10^{-3}$ M manganese(III) acetate initially filled the packed bed and a solution of 0.1 M CHD and $0.5 \times 10^{-3}$ M manganese(II) acetate was injected into the packed bed. This ensures a constant total concentration of Mn ions (Mn\textsuperscript{2+} and Mn\textsuperscript{3+}), and that changes in signal intensity come from changes in the oxidative state of the Mn\textsuperscript{3+}, rather than a depletion of manganese ions. All solutions were prepared in distilled, deionized water. These experiments were performed in high sulfuric acid conditions which stabilizes Mn\textsuperscript{3+} ions in aqueous solutions \cite{Cotton1999,Meyer1924}. We prepared fresh solutions before each experiment. The time scale of the reaction, which is visible optically, is calculated by a batch experiment. Using a CMOS camera, we first measure the time scale of dilution (without one of the reactants), and second the time scale of dilution+reaction. The difference between both times provides the reaction time scale, which is found to be $\tau_\textrm{r} \sim$ 1 s (see SI, Figure S2). 

MRI is the best technique to probe the concentration maps of manganese ions, because there is poor optical contrast between the two oxidative states, particularly in a packed bed.
\cite{Britton2006,Britton2017,Britton2010}. However, the Nuclear Magnetic Resonance (NMR) signal intensity ($I$) depends on concentration of manganese ions and their oxidative state (Mn\textsuperscript{2+} or Mn\textsuperscript{3+}) \cite{Britton2006}. There are more unpaired electrons in Mn\textsuperscript{2+} ions than Mn\textsuperscript{3+}, and as a result the relaxation time for molecules surrounding Mn\textsuperscript{2+} is shorter. This produces the necessary contrast with which to visualize reactive fronts inside the porous medium using MRI. Therefore, initially the non-wetting phase ($O$) and beads appear dark, and the resident Mn\textsuperscript{3+} ($B$) appears as the lighter phase. As the reaction occurs, the oxidative state of manganese ions changes which in turn changes the visible intensity. Note that the immiscible phase $O$ does not contain protons and hence is visualized by afterwards \cite{Ramskill2016}. The relaxation times (transverse or spin-spin relaxation times, $T_{2,m}$) for water can be converted into concentrations given that the total concentration of manganese ions remains constant ([Mn]\textsubscript{0} = [Mn\textsuperscript{2+}] + [Mn\textsuperscript{3+}]) \cite{Britton2006}.

\begin{equation}
\frac{1}{T_2} = k_\mathrm{x} %+ \frac{1}{T_{2,beads}}
\label{eqn:t2}
\end{equation}

\begin{equation}
\frac{1}{T_2} =  \frac{1}{T_{2,\mathrm{Mn^{2+}}}} + \frac{1}{T_{2,\mathrm{Mn^{3+}}}} + \frac{1}{T_{2,\mathrm{beads}}}
\label{eqn:t2_species}	
\end{equation}

\begin{figure*}
	\centering
	\includegraphics[width=16cm]{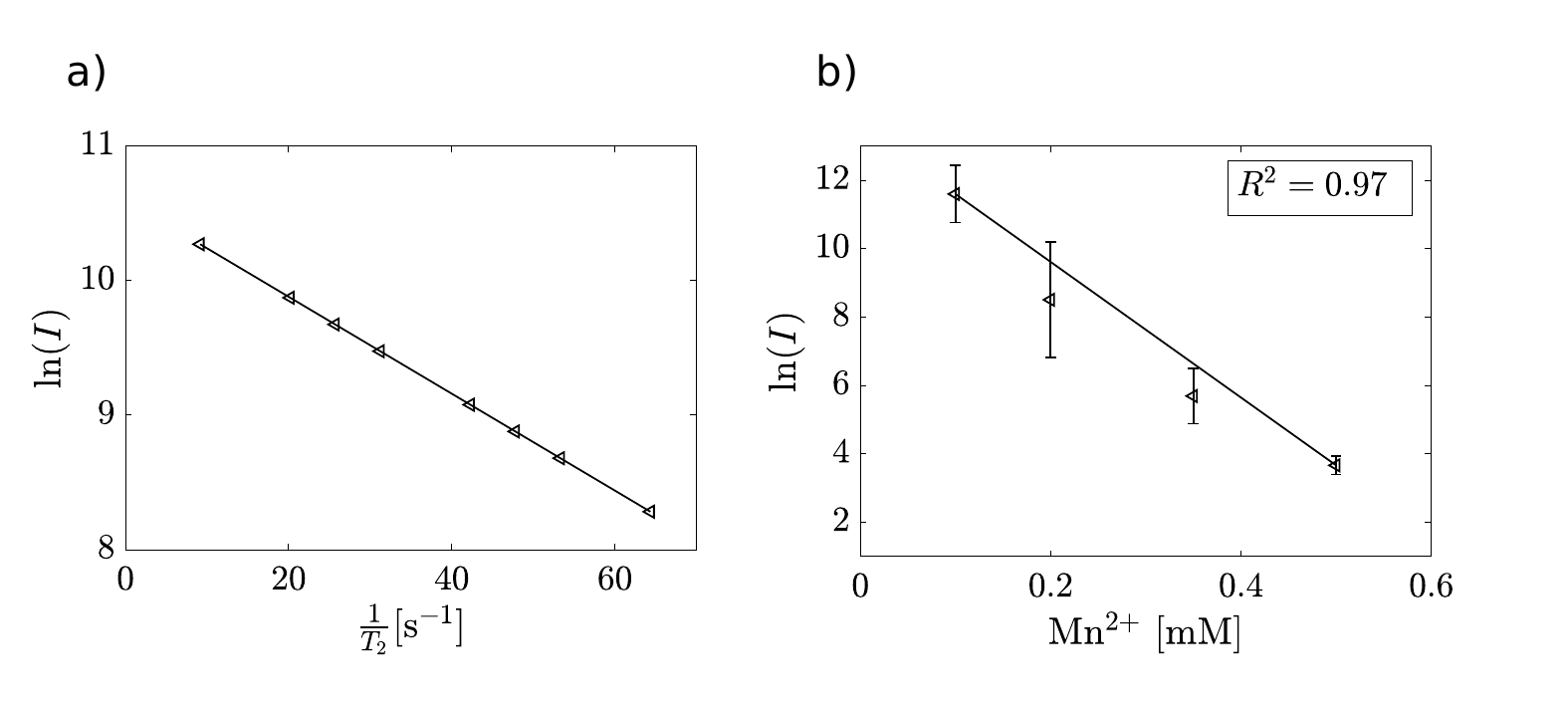}
	\caption[Calibration curve]{ a) The natural logarithm of the signal intensity ($I$) is proportional to the relaxation rate ($1/T_2$) (data from \cite{Britton2006}). b) Natural logarithm of $I$ against concentration of Mn\textsuperscript{2+} in the experiments.}
	\label{fig:calibration}
\end{figure*}

There is a simple linear relationship between the overall relaxation rate ($1/T_2$) of a paramagnetic species ($m$) and its concentration \cite{Britton2006_2} (Eq. \ref{eqn:t2}), however it is also affected by the diamagnetic relaxation of the solvent and proximity to the glass beads (see SI for details). The intensity observed is a combination of the total concentration of manganese ions [Mn] and the glass beads (Eq. \ref{eqn:t2_species}). Calibration experiments were performed in order to quantify the contribution of glass beads by only measuring the relaxation time (and hence the intensity) for each of the individual species ($m$). The natural logarithm of intensity ($I$) is directly proportional to the relaxation rate ($1/T_2$) \cite{Britton2006_2} (Figure \ref{fig:calibration} a). This is correlated to the concentration of species as shown in Figure \ref{fig:calibration} b. For our calibration, the data follows a linear fit with a regression coefficient of $R^2 = 0.97$. 

\subsection{Experimental Protocol}

The monodispersed 3$D$ porous media were used to run saturated and unsaturated experiments at different flow rates $Q$. For the unsaturated experiments, the immiscible phase ($O$) to 'desaturate' the system is injected from the upper side of the column, which was open (Figure \ref{fig:schematic}). A syringe to control the volume and a needle were used to allocate $O$ randomly (and homogeneously in statistical sense) within the porous medium. Once the beads and resident chemicals (either $B$, or $B$ and $O$ for saturated and unsaturated experiments, respectively) were placed inside the column, the latter was then carefully installed inside the magnet before the injection of $A$ (Figure \ref{fig:schematic}). Subsequently, the continuous injection of $A$ was started. While for the saturated experiments, and in both porous media, two different flow rates ($Q$) were used, the flow rate in the unsaturated ones was modified in order to get same Pe and very similar Da as in one of the saturated experiments for comparison. To ensure that there is no flow behavior missed between two consecutive scans, the upper limit of the imposed flow rate is determined by the acquisition time (16 s for a full 3$D$ scan), i.e., the acquisition time must be longer than the advective time ($\tau_\textrm{a}$). Owing to the capillary forces and low flow rates used, the immiscible phase was immobile during the course of the experiments, i.e., the magnitude of the viscous forces was smaller than the magnitude of capillary forces \cite{Tang2019}. The saturation degrees ($S_\textrm{w}$), the flow rates imposed, the resulting mean pore water velocities ($\overline{v}$), and the Péclet (Pe) and Damköhler (Da) numbers experimented are summarized in Table \ref{tab:exp_conditions}. 

1H and 19F magnetic resonance 3$D$ images were acquired using a Bruker Avance III HD spectrometer which comprised a 7 T wide-bore superconducting magnet operating at a proton resonance frequency of 300.13 MHz. All images were acquired using a micro 2.5 imaging probe equipped with a dual resonance 1H/19F 25 mm radio frequency (RF) birdcage coil. The temperature of the imaging probe was maintained at 293 ± 0.3 K by the temperature of water-cooled gradient coils. Vertical (sagittal) 3$D$ images of the system were acquired using the fast spin-echo imaging sequence RARE \cite{Rare1986}. 1H 3$D$ images were recorded using a $128\times 64\times 16$ pixel matrix, with a field of view of $40\times 20\times 20$ pixels. A RARE factor of 32, echo time of 3.2 ms and repetition time of 500 ms were used. For each experiment, a range of images (from 20 to 100) were collected depending on the injection time. 3$D$ 19F MR images were acquired to visualize the non-wetting phase $O$ at $t = 0$ (before the injection of $A$) and at the end of the injection \cite{Ramskill2016}. These 3$D$ images were acquired using the same parameters as the 3$D$ 1H MRI images, except for the repetition time, which was 1 s. Scans, at each time step, consisted of 16 slices (in the x direction), of 1 mm thickness, with a resolution of 0.25 mm (in the y and z directions). The size of each voxel is $0.25 \times 0.25\times 1$ mm, and thus the voxel volume equal to 0.0625 mm$^{3}$. Signal intensity was stored in 16-bit gray scale images. All images were processed and then converted to images of concentration, averaging over the voxel size, using the calibration described above (Figure \ref{fig:calibration}). Artifacts were corrected (see SI, Figure S3), and segmentation was used to differentiate the glass beads from the liquid phase. The segmented liquid volume is compared to the actual volume hosted in the porous medium. This action was also performed for the volume of the immiscible phase. The acquisition was carried out until no further reaction was detected in the visible domain. We compute for every time step the mass of Mn\textsuperscript{2+} injected and the excess mass visible in the pore volume as a consequence of reaction. The difference between these two provides the mass of Mn\textsuperscript{3+} that has reacted to produce Mn\textsuperscript{2+} in each voxel at every time step (see SI, Figure S4). We then calculate the total mass of product formed $M_P$ as the sum in all voxels in the visible domain. The rate of change of $M_P$ provides the effective reaction rate $R$.

\section{Results and Discussion} \label{section:results}

\subsection{Flow and Saturation Control on the Dynamics of Reaction}

\begin{figure}[h]
	\centering
	\includegraphics[width=8cm]{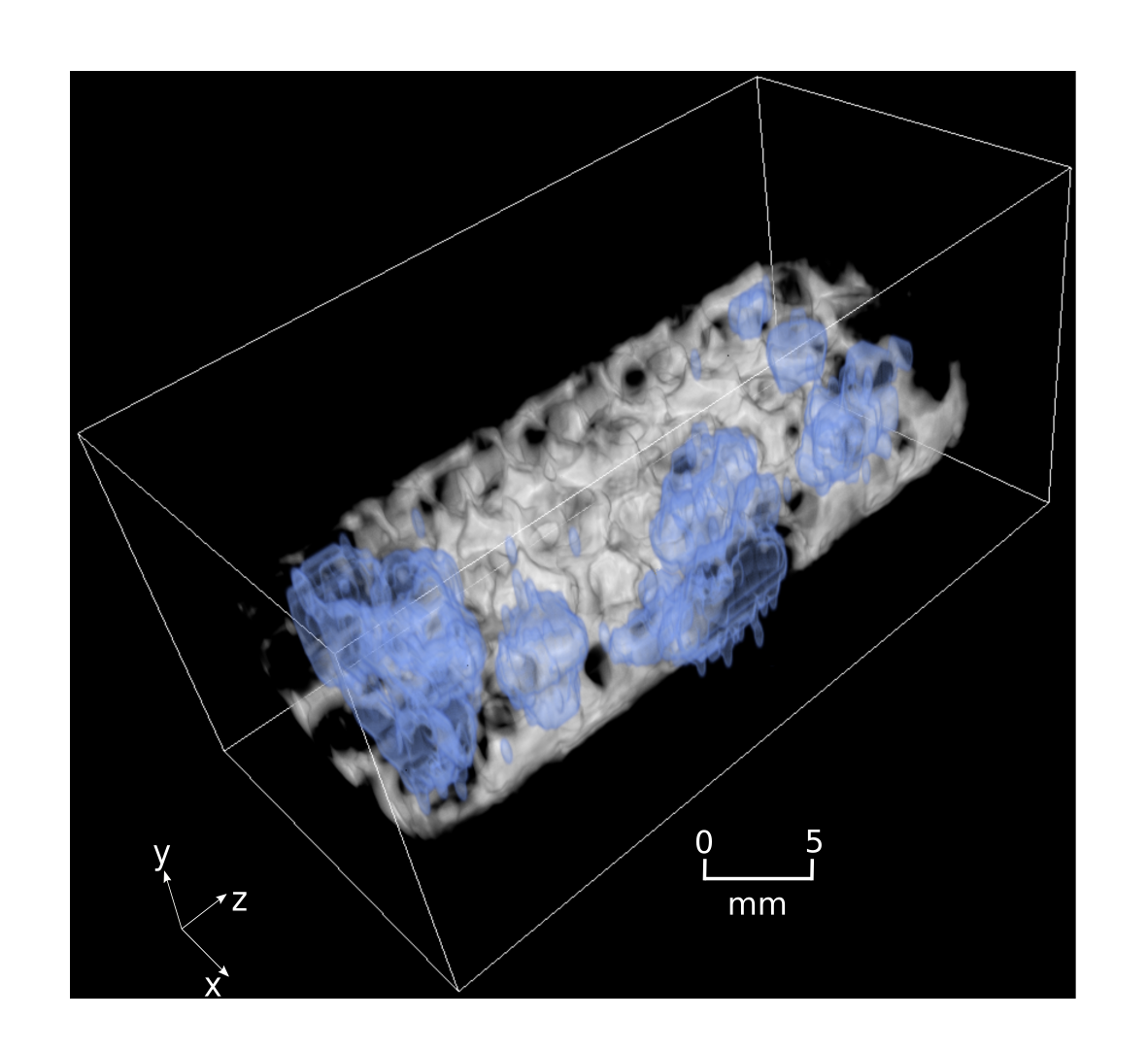}
	\caption{Distribution of the resident reactant $B$ (semi-transparent white color) and of the immiscible non-wetting phase $O$ (semi-transparent blue color) within the monodispersed porous medium of 4 mm glass beads (non visible). Saturation of the wetting phase is $S_\textrm{w}= 0.78$. $O$ is randomly distributed within the pore space.}
	\label{fig:phases_distribution}
\end{figure}

\begin{figure*}
	\centering
	\includegraphics[width=1\textwidth]{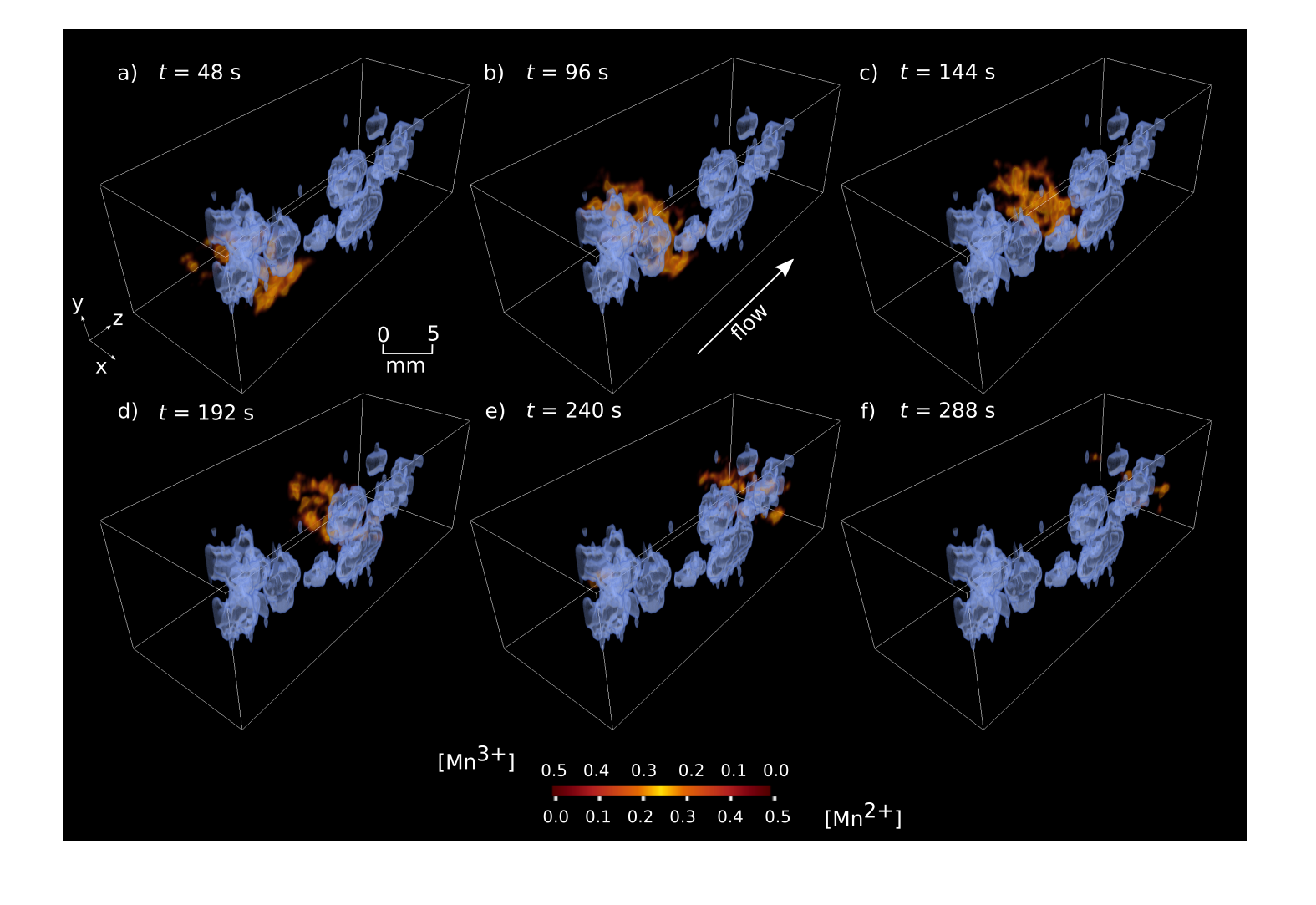}
	\caption{Time series of the transported reaction as $A$ is pumped at constant flow rate ($Q=7.697$ mm$^3$/s) into an unsaturated packed bed of 4 mm glass beads. Concentration of the reactants within the mixing volume is shown in warm colors, in which the lightest color indicates equal concentration of $A$ and $B$ (i.e., equal concentration of Mn\textsuperscript{2+} and Mn\textsuperscript{3+}, mM). Non-wetting phase $O$ is shown in a semi-transparent blue color and does not move during the experiment. Saturation of the wetting phase is $S_\textrm{w}= 0.78$. Pe and Da numbers for this experiment are 250 and 35, respectively. The arrow denotes the mean flow direction. Images were acquired every 16 s (see SI, Movie S2).} 
	\label{fig:4mm_timeSeries}
\end{figure*}

Figure \ref{fig:phases_distribution} shows the spatial distribution of the immiscible non-wetting phase $O$ in the pore space of a porous medium built from 4 mm glass beads. Clusters of this phase connecting several pores and isolated drops can be recognized. The saturation of the wetting phase in this case is $S_{\textrm{w}}=0.78$. $O$ was immobile during the reactive transport experiment. For the same experiment, snapshots at six equispaced times of the concentration of the reactants within the mixing volume are shown in Figure \ref{fig:4mm_timeSeries}, for an injection flow rate $Q=7.697$ mm$^3$/s. Experimental conditions of the reactive experiments in both porous media (i.e., different glass beads diameter) are summarized in Table \ref{tab:exp_conditions}. As the invading chemical $A$ enters the domain, it penetrates the channels created by the grains and $O$ (see Figure \ref{fig:4mm_timeSeries}). Initially the interface $\Pi$ is highly stretched and due to heterogeneity of the pore space a collection of lamellar topological structures (or fingers) develops (Figure \ref{fig:4mm_timeSeries} a-c). As the reaction propagates through the pore space, these fingers merge by diffusion and a more homogeneous reaction front propagates through the medium (Figure \ref{fig:4mm_timeSeries} d). After breakthrough, incomplete mixing makes the reaction persist behind the front (Figure \ref{fig:4mm_timeSeries} e and f) (see SI, Movie S2).

\begin{table*}
	\caption{Experimental conditions of the reactive experiments in the packed beds of 1 and 4 mm glass beads for the different saturation degrees $S_\textrm{w}$. $Q$ is the imposed flow rate, $\bar{v}$ is the mean pore water velocity, Pe is the Péclet number, and Da is the Damköhler number.}
	\label{tab:exp_conditions}
	\begin{tabular}{lllll|lllll}
		& & 1 mm grains & & & & & 4 mm grains&  &  \\
		\hline
		$S_\textrm{w}$ & $Q$ [mm\textsuperscript{3}/s] & $\bar{v}$ [mm/s]  & Pe & Da & $S_\textrm{w}$ & $Q$ [mm\textsuperscript{3}/s] & $\bar{v}$ [mm/s]  & Pe & Da \\
		\hline
		1.00 & 3.865 & 0.050  & 25 & 20 & 1.00 & 10.996 & 0.125  & 250 & 32  \\
		1.00 & 1.933 & 0.025  & 12.5 & 40 & 1.00 & 17.593 & 0.200  & 400 & 20 \\
		0.77 & 3.286 & 0.054  & 25 & 18 & 0.88 & 9.346 & 0.119  & 250 & 33\\
		0.43 & 1.933 & 0.058  & 25 & 17 & 0.78 & 7.697 & 0.112  & 250 & 35 \\
		\hline
	\end{tabular}
\end{table*}

%, with a decrease in its rate before breakthrough of the reactive front from our visible domain
%This is also reflected in the plot of $R$
%For low Pe (i.e., 1 mm grains), the rate of $M_P$ production after breakthrough is sustained longer in time under fully saturated conditions, although with a faster decay as compared to unsaturated conditions (Figure \ref{fig:R_Mp} c). 

As the chemicals react, we measure in time the effective reaction rate $R$ and the mass of reaction product $M_P$ for each experiment (symbols in Figure \ref{fig:R_Mp}). For all our experiments, $R$ initially increases. After breakthrough, $R$ rapidly decreases and eventually the reaction dies out. For a given saturation, in this case $S_\textrm{w} = 1$, a faster increase with time of $R$ as flow rate increases, i.e., as Pe increases, is observed, irrespective of the grain size. For a given Pe, a lower magnitude of $R$ as $S_\textrm{w}$ decreases is observed (Figure \ref{fig:R_Mp} a and b). The temporal evolution and magnitude of $M_P$ follows the patterns dictated by $R$, although some features are better recognized as follows. In Figure \ref{fig:R_Mp} c and d, it is seen that $M_P$ for fully saturated cases depends on the Pe. A higher Pe leads to a higher mass production. $M_P$ for unsaturated cases reduces as $S_\mathrm{w}$ decreases. However, for 1 mm sized grains (Figure \ref{fig:R_Mp} c), as $S_\textrm{w}$ reduces to 0.77, the mass produced at the very early times is larger than in fully saturated conditions. This trend reverses when $S_\mathrm{w}$ reduces to 0.43, i.e., the rate of increase of $M_P$ reduces as compared to $S_\mathrm{w}=0.77$. For low Pe (i.e., 1 mm sized grains), the rate of production of $M_P$ decreases before the breakthrough (denoted by vertical lines in Figure \ref{fig:R_Mp} c). On the contrary, for higher Pe (i.e., 4 mm sized grains), $M_P$ production does not decrease before breakthrough (Figure \ref{fig:R_Mp} d). After breakthrough, for both 1 and 4 mm sized grains, the rate of $M_P$ production decreases slower as saturation decreases (Figure \ref{fig:R_Mp} c and d). Note that despite the differences between the measured porosities (0.3844 and 0.4375, for 1 and 4 mm sized grains, respectively), the maximum mass produced for $S_\textrm{w} =1 $ in 1 mm sized grains and Pe = 25 is $M_P=43.4$ mg, whereas for the 4 mm sized grains and Pe = 250, it is $M_P=44.9$ mg. Thus the Pe and pore size plays a key role in determining $R$ and thus $M_P$. 

\subsection{Mixing Control on Reaction: Reactive Lamella Model Prediction}

\begin{table*}
	\caption{Parameters used in the model prediction. $s_0$ and $\varepsilon_0$ are the initial interface thickness and area, respectively. $\gamma$ is the shear deformation rate. $t_\mathrm{mix}^\mathrm{model}$ is the mixing time computed from the lamella based model and compared with the one inferred from the experiments. $t_\mathrm{mix}^\mathrm{exp}$ is experimentally observed only for 1 mm sized grains since it is not reached in the 4 mm sized grains experiments.}
	\label{tab:theo_params}
	\setlength\tabcolsep{5pt} % default value: 6pt
	\begin{tabular}{lllllll|lllll}
		& & 1 mm& grains & & & & & & 4 mm & grains & \\
		\hline
		$S_\textrm{w}$ & Pe & $\varepsilon_0 [\mathrm{m^2}]$ & $s_0 [\mathrm{m}]$& $\gamma [\mathrm{s^{-1}}]$& $t_\mathrm{mix}^\mathrm{exp} \mathrm{[s]}$ & $t_\mathrm{mix}^\mathrm{model} \mathrm{[s]}$ & $S_\textrm{w}$ & Pe & $\varepsilon_0 [\mathrm{m^2}] $ & $s_0 [\mathrm{m}] $& $\gamma [\mathrm{s^{-1}}]$ \\
		\hline
		1.00 & 12 & 0.0085 & 0.005 & 0.0626 & 208 & 185 & 1.00 & 400 & 0.0616 & 0.005 & 0.1251 \\
		1.00 & 25 & 0.0085 & 0.005 & 0.1251 & 128 & 116 & 1.00 & 250 & 0.0616 & 0.005 & 0.0782 \\
		0.77 & 25 & 0.0048 & 0.003 & 0.1059 & 110 & 92 & 0.88 & 250 & 0.0547 & 0.005 & 0.0556 \\
		0.43 & 25 & 0.0037 & 0.003 & 0.0624 & - & 132 & 0.78 & 250 & 0.0481 & 0.005 & 0.0461 \\
		\hline
	\end{tabular}
\end{table*}

\begin{figure*}
	\centering
	\includegraphics[width=1.0\textwidth]{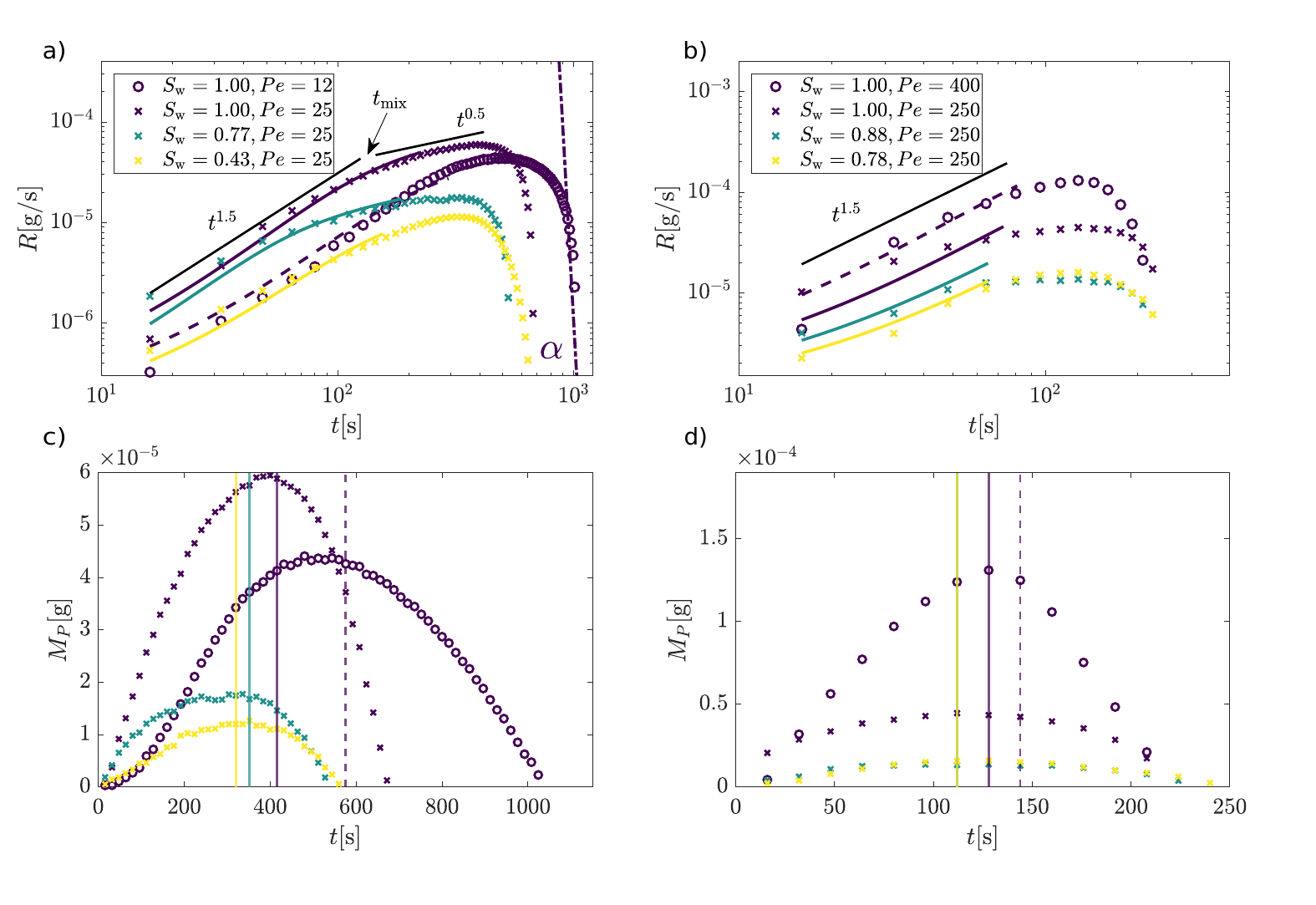}
	\caption{a, b) Comparison of temporal evolution of the global reaction rate $R$ for 1 and 4 mm sized grains, respectively, and different $S_\textrm{w}$ between the MRI experiment results (symbols) and the reactive lamella model (solid lines). Note that the same Pe is used for comparing different $S_\textrm{w}$ except where specified. The stretching ($t^{1.5}$) and the Fickian ($t^{0.5}$) regime are indicated with the solid black lines. The mixing time ($t_\mathrm{mix}$) between both regimes is also shown. The model prediction is shown until only breakthrough (i.e., the reaction front reaches the end of the observable domain). The dotted line in (a) depicts the speed $\alpha$ of reaction extinction after breakthrough. c, d) Temporal evolution of $M_P$ obtained from MRI experiments (symbols) for 1 and 4 mm sized beads, respectively, and different $S_\textrm{w}$. Vertical lines denote when breakthrough happens.}
	\label{fig:R_Mp}
\end{figure*}

We now compare the results obtained from the experiments with the theoretical model presented in Section \ref{section:theory}. We use the lamella model to describe the mixing interface as a collection of stretched lamellae and to predict the global reaction rate $R$ as function of $S_\textrm{w}$. The model parameters ($s_0$, $\varepsilon_0$, $\gamma$) used in Equation \ref{eqn:model} are given in Table \ref{tab:theo_params}. The solid lines in Figure \ref{fig:R_Mp} a and b show the results of the model (only until breakthrough) and how they compare with the experimental observations (symbols). 

At early times and for all $S_\textrm{w}$ and Pe, $R$ grows in time faster than Fickian, i.e.,  $R\sim t^{1.5}$. In the temporal evolution of $R$ for 1 mm sized grains, a change in the scaling from $t^{1.5}$ to $t^{0.5}$ for all $S_\textrm{w}$ before breakthrough is observed (Figure \ref{fig:R_Mp} a). This change happens at time $t_{\mathrm{mix}}$ when diffusion overcomes compression and concentration gradients are no longer enhanced. At later times, folding of the plume over itself promotes the lamellae interaction due to diffusion and they coalesce into bundles. \cite{LeBorgne2013}. For 4 mm sized grains, the front breaks through before $t_{\mathrm{mix}}$ and the coalescence regime, thus no change in the scaling of $R$ is observed (Figure \ref{fig:R_Mp} b). After $t_{\mathrm{mix}}$, $\varepsilon$ no longer grows linearly \cite{deAnna2014_2} and rate of product formation slows down. At longer times and for fully saturated conditions, $R$ is expected to decay in time as Fickian (i.e., $R \sim t^{-0.5}$) \cite{DeAnna2013}. However, for unsaturated conditions, this decay is expected to be slower than Fickian and it would reduce as $S_\mathrm{w}$ decreases, as observed for 2$D$ flows \cite{JJM2015,JJM2020}. Comparing different $S_\mathrm{w}$ (same Pe), $t_\mathrm{mix}$ is reached earlier at $S_\mathrm{w} =0.77$ than at $S_\mathrm{w} =1.00$, but a further reduction of $S_\mathrm{w}$ (i.e., $0.43$) increases the mixing time (Figure \ref{fig:R_Mp} a and Table \ref{tab:theo_params}). An analogy of this inversion in the trend can be found for conservative transport in unsaturated porous media, where a higher value of dispersivity and dispersion coefficient as saturation decreases is observed. \cite{muller2018influence,matsubayashi1997characteristics}  However, further observations indicate that this relation is not monotonic, and the maximum dispersivity and dispersion coefficient occurs at an intermediate saturation, called \textit{critical saturation}. \cite{Torideetal2003,RaoofHassanizadeh2013} While this has been explained by some authors for being the saturation where the tortuosity has its highest value, others argue the strong channeling effects in both fully saturated and low-saturation cases, being less significant at intermediate saturation values. \cite{birkholzer1997solute}

%For a higher Pe, $t_\mathrm{mix}$ is reached earlier (Figure \ref{fig:R_Mp} a).

\begin{figure*} 
	\centering
	\includegraphics[width=1.0\textwidth]{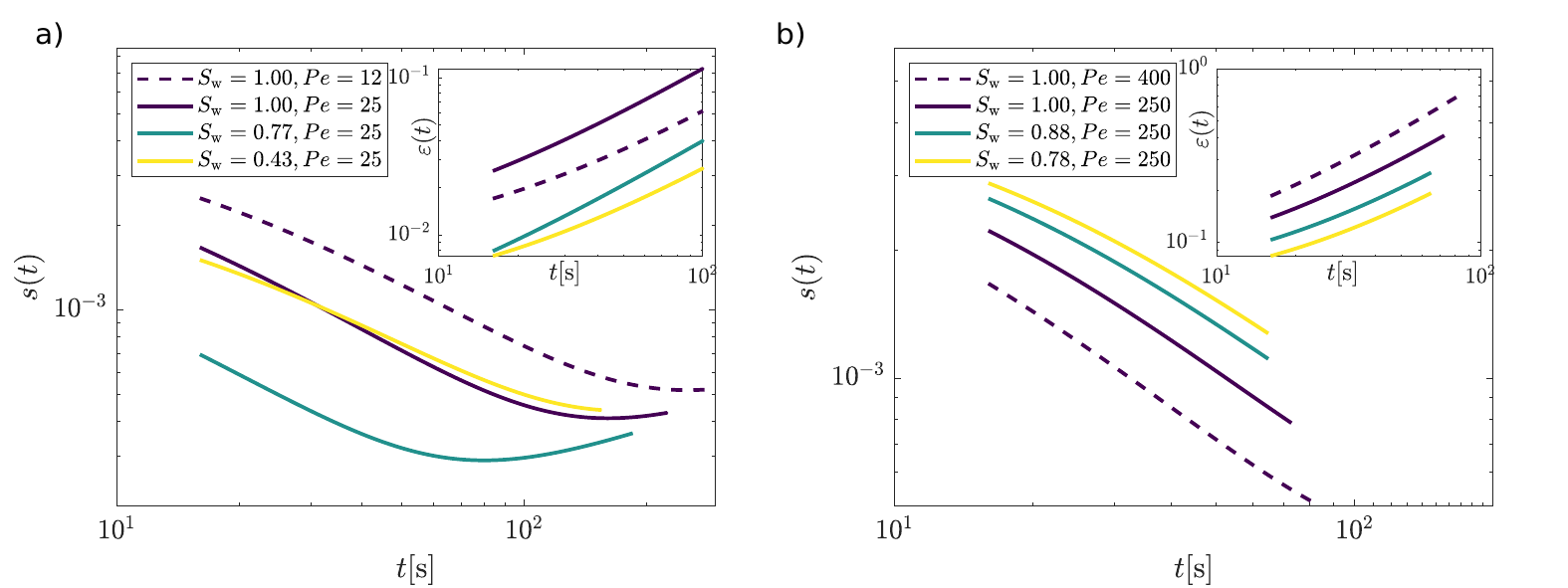}
	\caption{$s(t)$ and $\varepsilon$ (inset) evolution computed from the reactive lamella model ( (Eqs.~\ref{eqn:ADR_lamella}-\ref{eqn:s}) for the time experiment before breakthrough in a) 1 mm sized grains and b) 4 mm sized grains.}
	\label{fig:s_epsilon}
\end{figure*}

The model overall gives a good agreement to our observations (until breakthrough). Hence the assumption of linear stretching holds for spherical grains and the range of Pe used, based on the fact the shear is induced by the velocity gradients between the no-slip boundary condition at the grain walls and the maximum velocity at the pore center \cite{Rolle2019}. We hypothesize this assumption is also valid in unsaturated conditions as shown below, because the impact of the non-slip condition at the liquid-gas interfaces on transport processes has been recently demonstrated for being negligible.\cite{guedon2019pore, triadis2019anomalous} For a given Pe, as $S_\textrm{w}$ decreases, global reaction rate $R$ can scale slightly higher than $t^{1.5}$ (Figure \ref{fig:R_Mp} a). This can be explained by the temporal evolution of $s(t)$ and $\varepsilon$ before breakthrough (Figure \ref{fig:s_epsilon}). For 1 mm sized grains, $s(t)$ decreases by compression until it reaches a minimum after which it grows diffusively. $\varepsilon$ is highest for $S_\mathrm{w}=1$,  but a  slightly faster growth is observed for $S_\mathrm{w}<1$ due to a higher shear rate $\gamma$ (Figure \ref{fig:s_epsilon} a). For the range of Pe and $S_\mathrm{w}$ explored, a constant gradient of velocity results from shear deformation, which is characterized by the transient mixing front strictly increasing linearly in time even for low $S_\mathrm{w}$. According to \citeauthor{Rolle2019} \cite{Rolle2019}, a strong stretching regime (i.e., Pe'>Da') for the range of Pe and Da studied here is always experienced by the mixing and reactive front (Figure \ref{fig:s_epsilon}). We are able to characterize all the results (for different Pe and $S_\mathrm{w}$) using linear stretching ($\rho \sim t^1$). This also gives an insight into the permeability field of the domain: between a moderate and strong heterogeneity field \cite{LeBorgne2013}.

The reactive lamella model (Eq.~\ref{eqn:ADR_lamella}) reduces to three parameters, $s_0$, $\varepsilon_0$ and $\gamma$, which can be evaluated as $\bar{v}/(\xi/S_\mathrm{w})$. Based on these estimations and without any other further fitting, the model provides a reasonable estimate of the global reaction rate in both fully and partially saturated conditions over the range of Pe investigated by taking into account the incomplete mixing at the pore scale.

\subsection{Saturation Control on Reaction Extinction} \label{flush}

\begin{figure}
	\centering
	\includegraphics[width=9cm]{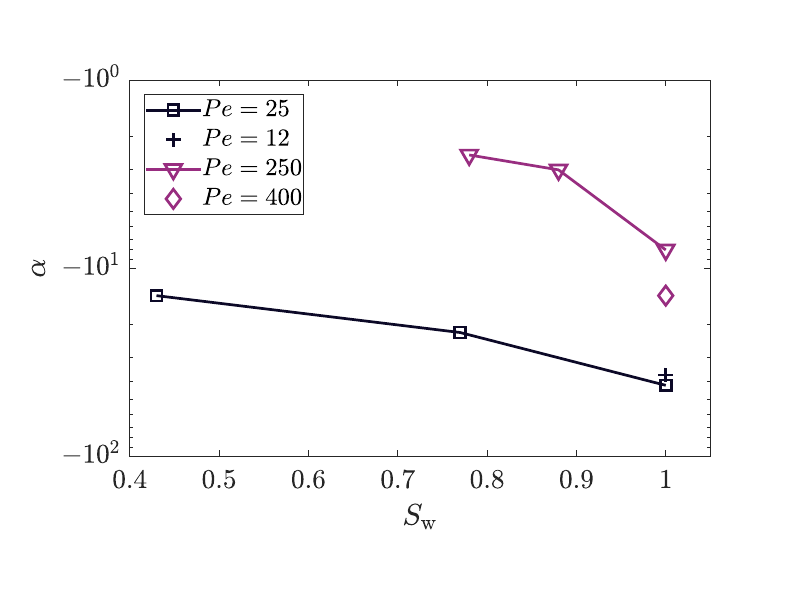}
	\caption{Impact of $S_\mathrm{w}$ on reaction extinction after breakthrough for different grain sizes and P: 4 mm sized grains, Pe = 250 and 400, purple symbols; 1 mm sized grains, Pe = 25 and 12, green symbols.}
	\label{fig:flush}
\end{figure}

Once breakthrough occurs, there is still reaction happening within the system due to incomplete mixing \cite{LeBorgne2011,Valocchi2018,Wright2017}. As saturation decreases, it takes longer for the reaction to completely extinguish (see Figure \ref{fig:R_Mp}). The presence of non-wetting phase $O$ increases the volume fraction of the wetting fluid with low velocities, i.e., stagnation zone. This promotes further reaction with the creation of concentration gradients between the high velocity regions or preferential paths and the low velocity regions. In some cases, the consumption of the resident reactant can take place only by diffusion \cite{JJM2016,JJM2017}. We define $\alpha$ as the rate of reaction extinction. As depicted in Figure \ref{fig:R_Mp}, $\alpha$ is the slope of the reaction rate $R$ tail calculated after breakthrough has already taken place. Figure \ref{fig:flush} compares $\alpha$ for both grain sizes and all Pe studied. $\alpha$ increases as $S_\mathrm{w}$ decreases. While previous 2$D$ observations shown an enhancement of mixing and reaction as $S_\mathrm{w}$ decreases \cite{JJM2016,JJM2017}, this is exacerbated in 3$D$ by the presence of helical flow components and transverse mixing \cite{Cirpka2014,Ye2016,Ye2018}. For $S_\mathrm{w} = 1$, increasing Pe (more than an order of magnitude, from 25 to 250) does not increase $\alpha$ by the same amount (Figure \ref{fig:flush}). When $S_\mathrm{w} < 1$, $\alpha$ increases by a similar order of magnitude independently of the Pe. Thus, a higher Pe does not necessarily lead to an increase in resilience of the reaction, it is in fact $S_\mathrm{w}$ which is the main controller of how long a reaction lasts inside the pore space. 

We have demonstrated and quantified experimentally the incomplete mixing at pore scale in 3$D$ porous media and the impact of the presence of an immiscible phase on it. The mixing and reactive front have been depicted by a lamella like topology. The front is advected and deformed by the heterogeneous velocity field, being subjected to stretching and folding and resulting in a competition between compression (creation of concentration gradients) and diffusion (destruction of concentration gradients). Compared to the 2$D$ case, we get a more stretched 3$D$ reaction front, however, the mechanisms which control the dynamics remain same. Although the fringes of the plumes are considered reaction hotspots \cite{McClain2003,Stegen2016,Pool2018}, subsurface environments are in general poorly mixed \cite{Kitanidis2012}. For instance, the initial response (at early times) can be crucial in understanding biogeochemical processes in unsaturated soils. While we considered a fast reaction compared to the transport in a small domain, such reactions are relatively common in natural environments \cite{sparks2013kinetics}. These findings have implications for effective reactive transport modeling in a variety of applications, since the basic phenomena we studied occurs in a wide range of flows.

\begin{acknowledgement}

IM and JJM acknowledge the financial support from the Swiss National Science Foundation (SNF, grant Nr. 200021\_178986). MMB and GC acknowledge the financial support from the University of Birmingham and the Engineering and Physical Science (EPSRC) Research Council, UK (EP/K039245/1).

\end{acknowledgement}

\begin{suppinfo}

The following files are available as Supporting Information (SI).
\begin{itemize}
  \item Movie\_S1.mp4: Propagation of reaction front inside the pore space of 8 mm sized glass beads.
  \item Movie\_S2.mp4: Propagation of reaction front (and incomplete mixing after breakthrough) in the pore space of an unsaturated 4 mm sized glass beads.
  \item SI.pdf: Reaction time scale from batch experiment, CFD simulation of column design and details on calibration procedure.
\end{itemize}

\end{suppinfo}

\bibliography{MRI_reactions}

\end{document}

\maketitle

\section*{Contents of this file}

\begin{itemize}
	
	\item Column design
	\item Reaction time scale calculation
	\item Calibration 
	
\end{itemize}

\section*{Additional Supporting Information (files uploaded separately)}

\begin{itemize}
	
	\item Captions for movies S1 and S2
	
\end{itemize}

\section{Column Design}

\begin{figure}[h]
	\centering
	\includegraphics[width=12cm]{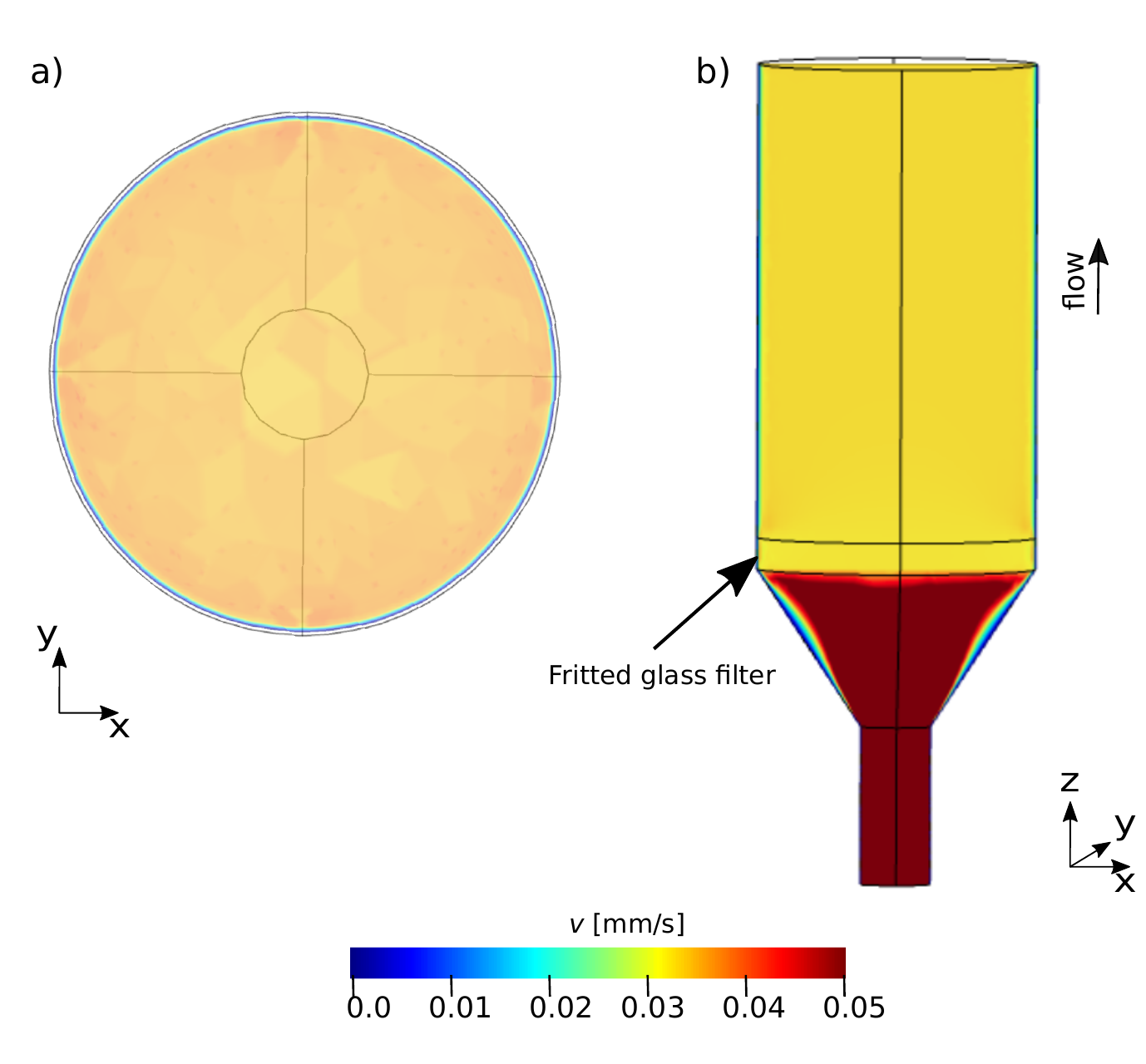}
	\caption*{Figure S1: a) 2$D$ section of steady state velocity field after passing the fritted glass filter perpendicular to the flow direction for a constant flow rate  $Q$=20 mm\textsuperscript{3}/s from bottom to top. b) Velocity field in the flow direction along a vertical cross-section in the middle of the domain. }
	\label{fig:cone}
\end{figure}

A Darcy-Brinkman formulation is used to simulate the fluid flow velocity in the conical shape of the vessel, in the fritted glass filter (2 mm thickness; $\sim$ 500 µm pore size) and in the porous medium (1 and 4 mm sized glass beads). While Navier-Stokes equations are used in the first case, a Darcy (continuum) formulation is used in the two porous media. Figure S1a shows how the fritted glass filter homogenizes the flow before the porous medium of interest and reaction occurs. Figure S1b shows the velocity field in the flow direction along a vertical cross-section. The velocity profile shows that for the given length of the cone, the flow remains attached to the sidewalls for a flowrate $Q$=20 mm\textsuperscript{3}/s (which is the highest flowrate used in the experiments). 
%Thus, our experiments can be considered in a steady state regime.

\section{Reaction time scale calculation}

\begin{figure}[h]
	\centering
	\includegraphics[width=12cm]{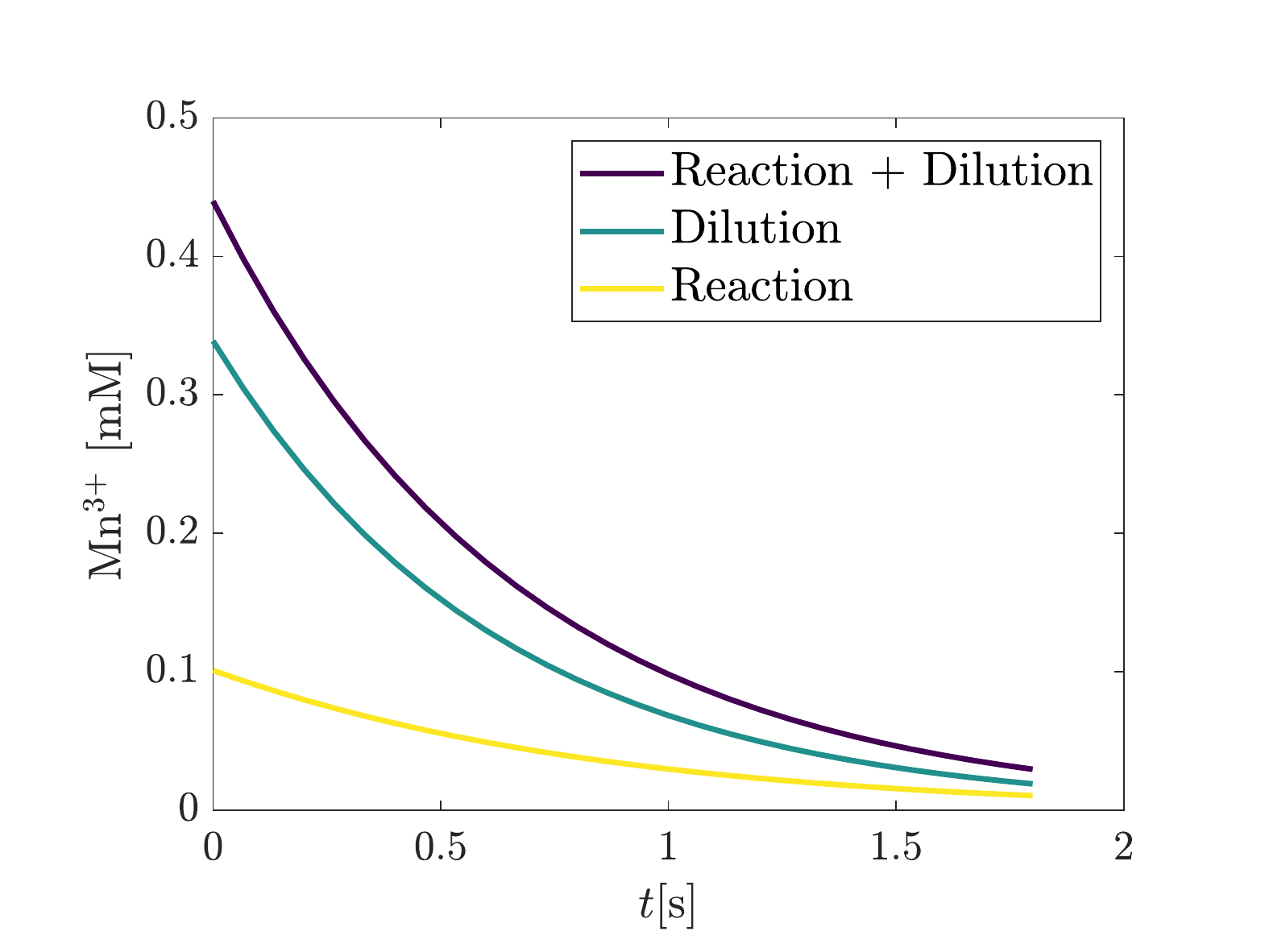}
	\caption*{Figure S2: Exponential curve fitted to the reaction and dilution processes when Mn\textsuperscript{3+} to Mn\textsuperscript{2+} reduction reaction occurs.}
	\label{fig:kinetics}
\end{figure}

A manganese catalysed 1,4 cylohexanedione (CHD) acid reaction is used to study the reaction kinetics (Eq. 5). Mn\textsuperscript{3+} is first produced by adding manganese (III) acetate to a strong acid H\textsubscript{2}SO\textsubscript{4} (2.5 M). The reaction which consists of conversion between the two oxidative states of manganese from Mn\textsuperscript{3+} to Mn\textsuperscript{2+} is accompanied by a change in colour. We captured this colour change using a CMOS camera at 15 frames per second. The visible process in the camera is that of reaction and dilution. The dilution was captured in another experiment where there is no reaction occurring. To determine the rate constant of the reaction, the effect of dilution has to be removed (green curve in Figure S2). Fitting the reaction curve to an exponentially decaying function, a $1^{\textrm{st}}$ order reaction with a rate constant $k=1.3$ is determined. Thus, the time scale of the reaction is $1/k\approx1$ s.

\section{Calibration}

\begin{figure}[h]
	\centering
	\includegraphics[width=8cm]{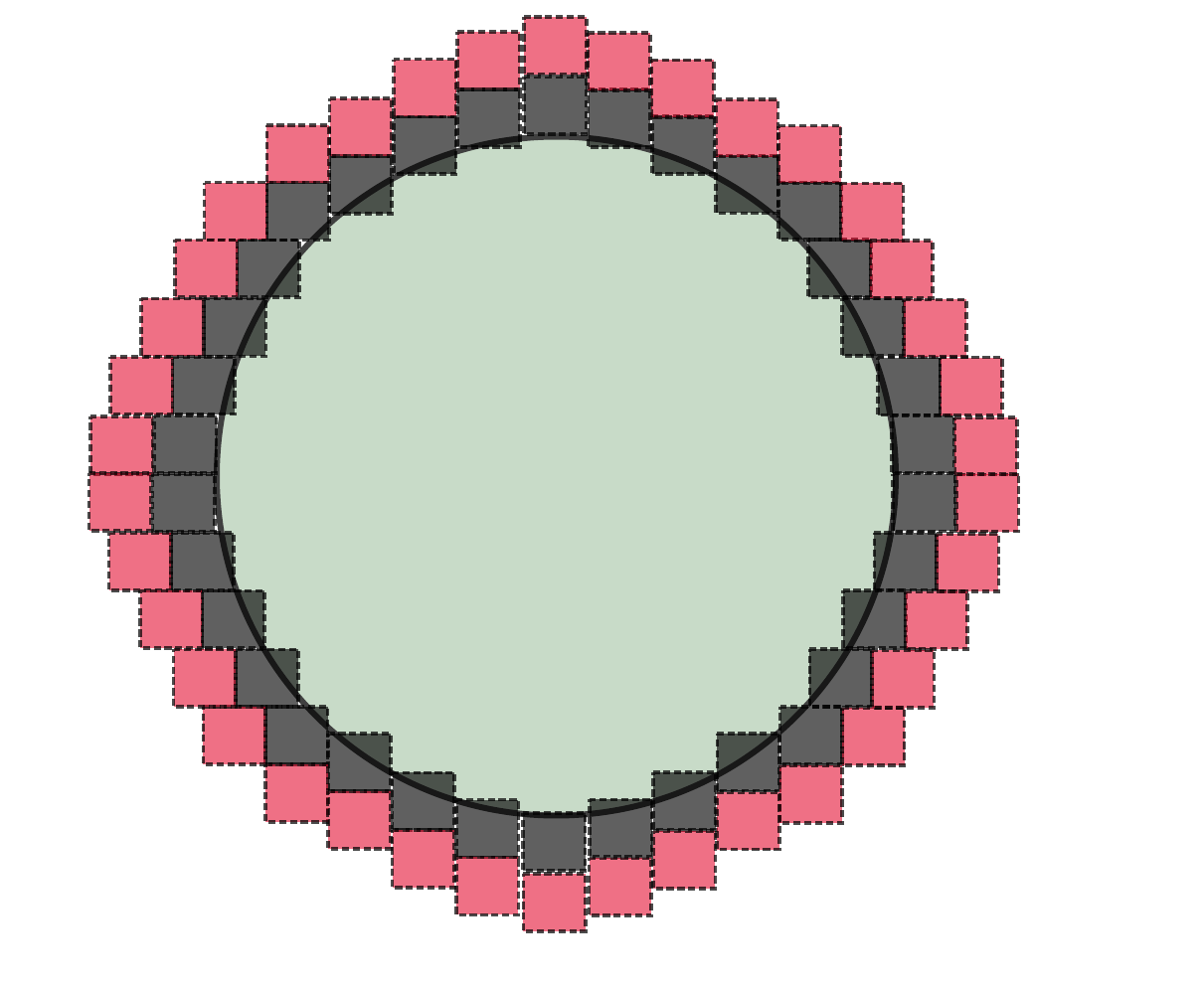}
	\caption*{Figure S3: Schematic showing MRI artifacts close to the beads. A representative glass bead is shown in green colour, the voxels with artifacts in gray and voxels without artifacts in pink.}
	\label{fig:artifacts}
\end{figure}

Before injecting any reactant, we measure the signal intensity in each voxel. Due to the difference in magnetic susceptibility differences of the beads and liquid, there exist artifacts very close to the glass beads. In Figure S3, as an example we show that gray voxels next to the beads have these artifacts which gives an incorrect concentration map based on the intensity observed. For the pink colored voxels there exist no artifacts. At initial time, all the liquid is unreacted solvent (and at final time all liquid is reacted solvent), hence the gray voxels must have the same intensity as the pink ones. For intermediate images, as the reaction occurs, we correct the intensity values for the gray voxels. We observe that the temporal changes in intensities for all the pink voxels follow a normal distribution. We use the mean value of this distribution to predict the intermediate intensity values in the gray voxels. Our method is corroborated by the fact that we see mass conservation in the domain, i.e.. the mass produced (of Mn\textsuperscript{2+}) in the pore space is same as what is calculated theoretically. This is further explained by Figure S4.

\begin{figure}[h]
	\centering
	\includegraphics[width=12cm]{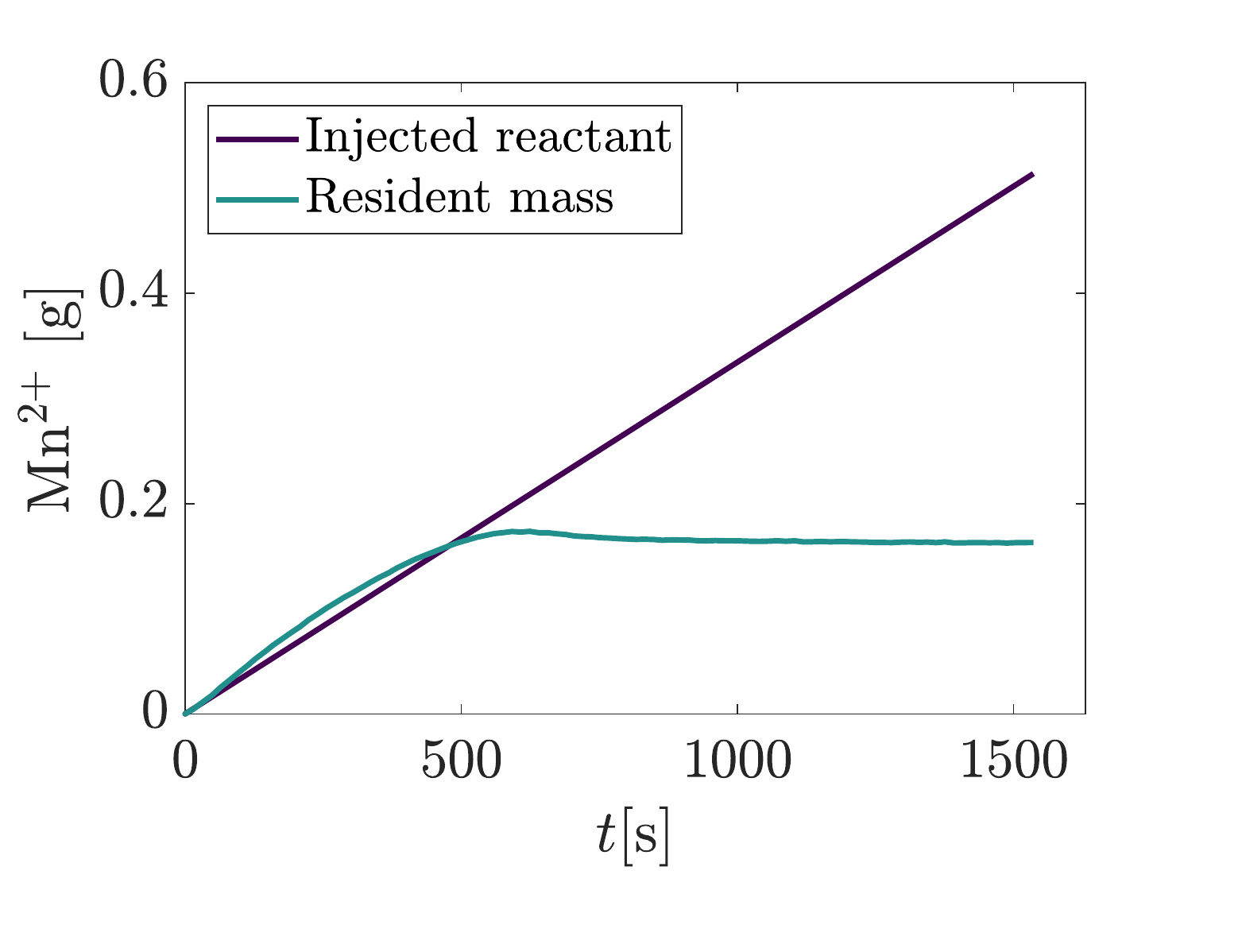}
	\caption*{Figure S4: Mass of Mn\textsuperscript{2+} plotted as a function of time.}
	\label{fig:artifacts}
\end{figure}

As an example, Figure S4 shows in green colour the resident mass of the reaction product within the observable domain, i.e., injected Mn\textsuperscript{2+} plus the mass produced by the reaction, as a function of time for 1 mm sized glass beads with $S_\mathrm{w}=1$ and Pe = 25. The mass of reaction product is calculated as the difference between the resident mass and the injected mass. The final resident mass can be calculated as following $M_\mathrm{Mn^{2+}}=c\cdot M_\mathrm{w}\cdot \phi \cdot V = 0.174$ g where $c$ = 0.5 mM is the concentration, $M_\mathrm{w} = 173.027$ g/mol is the molecular weight of Mn\textsuperscript{2+}, $\phi = 0.38$ is the porosity of the system (1 mm beads) and $V$ = 5286 mm\textsuperscript{3} is the volume of the observed region. The final resident mass is used to validate the one inferred from MRI. 

\section*{Caption Movie S1}

Transport of the reactive front inside a fully saturated ($S_{\textrm{w}}=1$) packed bed of 8 mm glass beads. Concentration of the reactants within the mixing volume is shown in warm colors, in which the lightest color indicates equal concentration of the invading ($A$) and resident ($B$) reactant. $A$ is pumped into the porous medium at constant flow rate ($Q=0.252$ mm$^3$/s) from bottom to top. Note that the lamella shown is cut by an x-z (vertical) plane passing through the middle of the domain; beads (in grey color) are not cut by the plane. Only three beads are shown for illustration purpose. 

\section*{Caption Movie S2}

Transport of the reaction as $A$ is pumped at constant flow rate ($Q=7.697$ mm$^3$/s) into an unsaturated packed bed of 4 mm glass beads. The flow is from left to right. Concentration of the reactants within the mixing volume is shown in warm colors, in which the lightest color indicates equal concentration of $A$ and $B$ (i.e., equal concentration of Mn\textsuperscript{2+} and Mn\textsuperscript{3+}, mM). Non-wetting phase $O$ is shown in a semi-transparent blue color and does not move during the experiment. Saturation of the wetting phase is $S_\textrm{w}= 0.78$. Pe and Da numbers for this experiment are 250 and 35, respectively. Reaction hotspots are visible in certain locations even after breakthrough has occurred. Note that the beads are not shown in this case.